\documentclass[a4paper]{PoS}

\usepackage{mathtools}

\newcommand{\g}{\gamma}
\newcommand{\e}{\epsilon}
\newcommand{\G}{\Gamma}

\title{Analyzing the December 2013 Orphan Gamma-Ray Flare From 3C 279}

\ShortTitle{Analyzing a 3C 279 Orphan Gamma-Ray Flare}

\author{\speaker{Tiffany R. Lewis}
	\thanks{This presentation/publication has received funding from the European Union’s Horizon 2020 research and innovation programme under grant agreement No 730562 [RadioNet].}
	\thanks{This work was supported in part by the Zuckerman STEM Leadership Program}\\
	Zuckerman Postdoctoral Scholar\\
        Center for Theoretical Physics \& Astrophysics, University of Haifa\\
        Department of Astrophysics, Tel Aviv University\\
        E-mail: \email{tiffanylewisphd@gmail.com}}

\author{Justin D. Finke\\
        U.S. Naval Research Laboratory,  Code 7653 \\
        4555 Overlook Ave. SW, Washington, DC, 20375-5352\\
         }
       
\author{Peter A. Becker\\
        Department of Physics \& Astronomy, George Mason University\\
        4400 University Drive, MSN 3F3, Fairfax, VA 22030
         }

\abstract{Multiwavelength monitoring of the blazar 3C 279 observed a very bright, 12-hour, orphan gamma-ray flare on 20 Dec 2013 with a uniquely hard {\it Fermi}-LAT spectrum and high Compton dominance. We work with a one-zone, leptonic model with both first- and second-order Fermi acceleration, which now reproduces the unique flaring behavior.  We present a simplified analytic electron energy distribution to provide intuition about how particle acceleration shapes multi-wavelength blazar jet emission spectra. The contributions of individual processes in relativistic jets is fundamental to understanding the particle energy budget in the formation and propagation of astrophysical jets. We show that first- and second-order Fermi acceleration are sufficient to explain the flare, and that magnetic reconnection is not needed. Our analysis suggests that the flare is initiated by an increase in the particle energies due to shock acceleration, which also increases the stochastic acceleration.  The higher energy particle preferentially occupy the outer jet, along the sheath, which decreases the apparent magnetic field and synchrotron radiation, while increasing electron exposure to the broad line region photon fields, driving up the external Compton emission. }

\FullConference{High Energy Phenomena in Relativistic Outflows VII - HEPRO VII\\
		9-12 July 2019\\
		Facultat de Física, Universitat de Barcelona, Spain}

\begin{document}

\section{Motivation \& Methodology}

Multi-wavelength monitoring of the blazar 3C 279 observed a very bright, 12-hour, orphan gamma-ray flare on 20 Dec 2013 with a uniquely hard, {\it Fermi}-LAT spectrum and high Compton dominance \cite{hayash15}.  An orphan flare may occur in any broad band if a flare is not observed simultaneously in another.  The driving question was how the blazar jet emitting region could produce such a powerful flare with a flux-doubling timescale of only $\sim2$ hrs.  To investigate this question, it is necessary to look at the acceleration mechanisms likely to act in a blazar jet.  Blazar jets are expected to accelerate particles via first- and second-order Fermi acceleration, which were treated directly, and magnetic reconnection, which was examined but not directly modeled. 

A one-zone, leptonic particle transport equation in the Fokker-Planck formalism allows for the examination of individual acceleration mechanisms on the electron number distribution $N_e(\g)$ with respect to the electron Lorentz factor $\g$.  The particle transport model utilized here includes first-order Fermi acceleration and second-order Fermi acceleration in addition to synchrotron and full Compton losses.  We assume that primary acceleration and emission zones are co-spatial and that particles enter this region with a thermal distribution that can be approximated by a delta distribution.  The steady-state particle transport equation \cite{lewis18}
\begin{equation}
0 = \dot{N}_{\rm inj} \delta(\gamma-\gamma_{\rm inj})  - \frac{N_e D_0 \g}{\tau}   
+ \frac{\partial^2}{\partial \gamma^2} \left( D_0\g^2 N_e \right)
- \frac{\partial}{\partial \gamma} \left( \left< \frac{d \gamma}{d t} \right> N_e\right)  \ ,
\label{eq-transport}
\end{equation}
includes terms from left to right for electron injection (at rate $\dot{N}_{\rm inj}$ and energy $\g_{\rm inj}$), energy-dependent electron escape (with timescale $\tau/(D_0 \g)$, the broadening term component of second-order Fermi acceleration, and the drift term where the coefficient \cite{lewis18}
\begin{equation}
\left< \frac{d \gamma}{d t} \right> = 
D_0 \left[a\gamma +4\gamma-b_{\rm syn}\gamma^2 - 
\gamma^2 \sum_{j=1}^J b^{(j)}_{\rm C} H(\gamma \epsilon_{\rm ph}^{(j)}) \right] \ ,
\label{eq-driftKN}
\end{equation}
includes the primary mechanisms through which particles gain and lose energy.  

The first term in the drift coefficient is the first-order acceleration term whose strength is determined by the parameter $a = A_0/D_0$ ($D_0$ appears throughout the equation due to the analytic solution method \cite{lewis16}). First order Fermi acceleration occurs as the particles in the acceleration region encounter the face of a compressive shock and gain energy in proportion to the energy they already have.  Thus, first-order acceleration is more efficient for more energetic particles. Charged particles preferentially follow magnetic field lines, and blazar emission is partially polarized, indicating that the magnetic field is turbulent or tangled. Thus, some particles scatter through many times, creating even higher particle energies with increased efficiency. Adiabatic expansion is included in this term because neither first-order acceleration nor adiabatic expansion appears anywhere else in the model.  Thus, acceleration contributes a positive value and adiabatic expansion contributes a negative value, but pragmatically only their sum is apparent.  

The second term is the drift coefficient component of second-order Fermi acceleration, which is tuned with the parameter $D_0$.  Second-order Fermi acceleration occurs as the particle population encounters magnetohydrodynamic scattering centers, which move stochastically in the co-moving frame. Particles tend to gain energy on average from these interactions due to the “head-on” advantage in a relativistically propagating region.  The rate of energy electrons gain by stochastic scatterings is controlled by the corresponding parameter $D_0$.  (See \cite{lewis16} for further details on the model parameters and their derivation.)

The third term in the drift coefficient represents energy that electrons lose to synchrotron radiation, and is parameterized by $b_{\rm syn}$ which depends on the magnetic field strength \cite{lewis16}.  The fourth term in the drift coefficient represents energy that electrons lose to inverse Compton scattering, which is mainly parameterized by $b_{\rm C}^{(j)}$, which depends on the energy density of the incident photon field $j$.  The model includes as incident photon fields, the accretion disk, dust torus, and 25 broad emission lines individually, which is why this term sums over the incident photon fields.  The function $H(\g\e)=1$ in the Thomson approximation, and otherwise includes the Klein-Nishina effect in the energy loss rate, where $\g$ is the electron Lorentz factor and $\e$ is the incident photon energy - each has a distribution. (See \cite{lewis18} for further discussion of the energy loss mechanisms.)

There is an analytic solution to the electron transport equation in the Thomson approximation for which the synchrotron and Thomson losses can be combined ($b = b_{\rm syn} + b_{\rm C}$) \cite{lewis16}
\begin{equation}
N_e(\g) = \frac{\dot{N}_{\rm inj}}{bD_0} \frac{\G(\mu - \kappa + 1/2)}{\G(1+2\mu)} e^{(\g_{\rm inj}-\g) b/2} \g_{\rm inj}^{-2-a/2} \g^{\, a/2} 
\begin{cases}
{\mathcal M}_{\kappa,\mu}(b \g) {\mathcal W}_{\kappa, \mu} (b\g_{\rm inj}) \quad , \g \le \g_{\rm inj} \\
{\mathcal M}_{\kappa,\mu}(b \g_{\rm inj}) {\mathcal W}_{\kappa, \mu} (b\g)  \quad , \g_{\rm inj} \le \g
\end{cases} \ .
\label{eq-analytic}
\end{equation}
The analytic solution includes $\G$-functions and Whittaker-functions ($\mathcal M, \mathcal W$) parameterized by the Whitaker coefficients ($\kappa$, $\mu$), which depend on $a, \, b$, and $\tau$.  (See \cite{lewis16} for further details.) The full Klein-Nishina inclusive equation can be solved numerically, but both electron distributions have similar properties with regard to the effect of acceleration mechanisms on the electron energy distribution. Thus, the analytic solution can inform the numerical results. This allows us to examine the acceleration mechanisms and driving energetics in flat spectrum radio quasars (FSRQ).  

\section{Particle Acceleration in Spectral Energy Distribution Modeling}

Each component of the broadband spectrum in an FSRQ is calculated from the particle energy distribution.  The electrons emit synchrotron radiation, and the shape of the synchrotron spectrum depends on the distribution of energy in the emitting electrons.  Similarly, the synchrotron self-Compton (SSC) and external Compton (EC) processes depend on the electron energies.  Different types of acceleration give the electron energy distribution different forms.  First-order Fermi acceleration alone gives a power-law electron distribution \cite{fermi49}, while second-order Fermi acceleration alone gives a log-parabola \cite{tramacere11}.  

\begin{figure}
\centering
       \includegraphics[width=.6\textwidth]{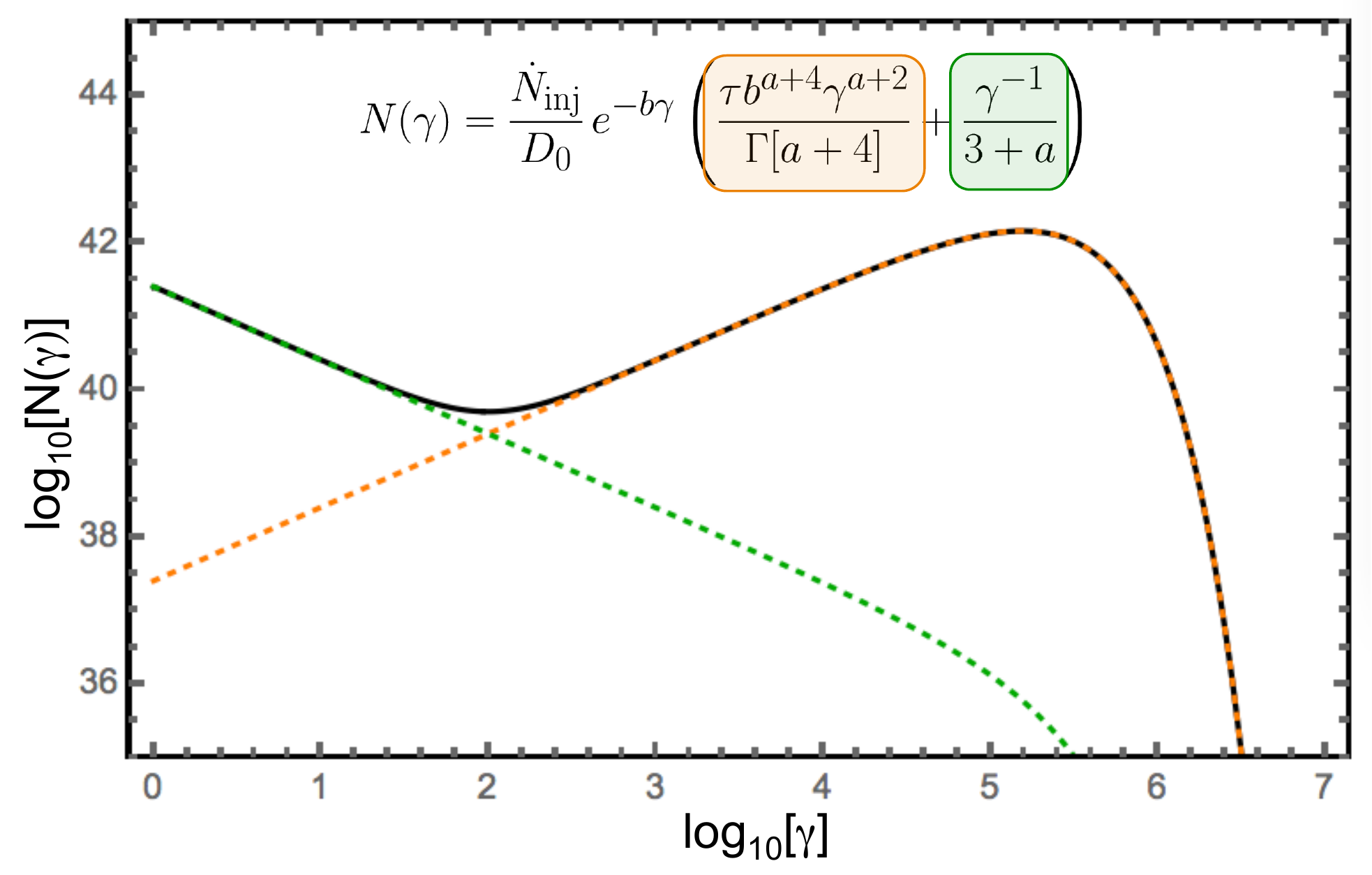}
       \caption{A sample electron energy distribution with respect to the electron Lorentz factor is shown with unphysical, exaggerated parameters to better illustrate the components.  The simplified version of the analytic solution to the Thomson transport equation helps to intuit how the acceleration mechanisms influence the electron energy distribution.  The green curve is due to stochastic acceleration while the orange is due to a balance between both of the acceleration terms (including adiabatic expansion). Figure published in \cite{lewis19}. }
       \label{fig-SimpED}
     \end{figure}

When both first- and second-order acceleration are included in the combined shock and stochastic acceleration produces a broken power law, although the break is not always apparent in practice if one term is categorically dominant.  This is best illustrated for the subset of the parameter space, where the solution can be expressed in a significantly simplified form, as shown in Figure \ref{fig-SimpED} \cite{lewis19}. There, the acceleration mechanisms clearly contribute different components to the electron energy.  In most cases of spectral energy distribution data analysis, the main difference is that the orange component will have a negative (rather than a positive as depicted) slope, due to a lower first- to second-order acceleration ratio.

\begin{figure}
\centering
       \includegraphics[width=.6\textwidth]{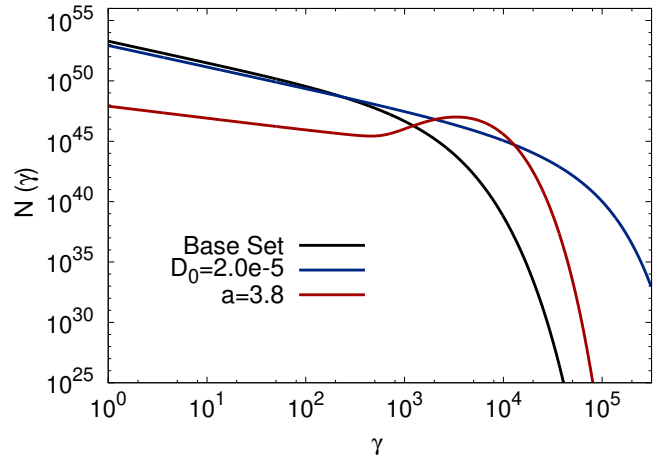}
       \caption{Electron energy distributions from the parameter study shown in Figure \ref{fig-SEDacc} are shown with respect to the electron Lorentz factor.  The black electron distribution has relatively low levels of both accelerations.  The blue distribution has higher levels of second-order Fermi acceleration, while the red has higher levels of first order Fermi acceleration.  Figure published in \cite{lewis18}.}
       \label{fig-EDacc}
     \end{figure}
     
The electron distributions in Figure \ref{fig-EDacc} were produced using the numeric solution to Equation \ref{eq-transport}.  In the blue curve, where the second-order acceleration was increased, the balanced term (orange from Figure \ref{fig-SimpED}) has a positive slope and dominates out to higher energies until the loss mechanisms give an exponential cutoff.  In the red curve of Figure \ref{fig-EDacc}, the second order term (green from Figure \ref{fig-SimpED}) is visible at lower energies and meets with the combined term at $\g \approx 10^3$, where the shock  causes a build-up of higher Lorentz factor particles. 

\begin{figure}
	\centering
       \includegraphics[width=.9\textwidth]{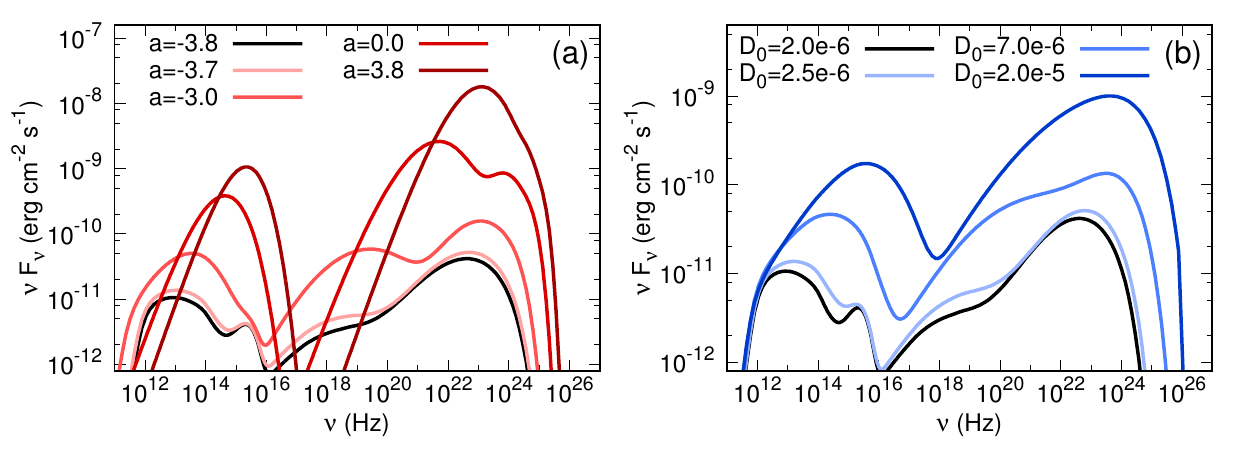}
       \caption{Multiwavelength spectral energy distributions from a parameter study are shown with respect to frequency. The color scheme is consistent with Figure \ref{fig-EDacc}. The black spectral curves in either frame are the same, and have relatively low levels of both accelerations.  In frame (a), the first-order Fermi acceleration is greater for more intense shades of red.  In frame (b), the second-order Fermi acceleration is greater for more intense shades of blue.  Figure published in \cite{lewis18}. }
       \label{fig-SEDacc}
     \end{figure}

The spectra that correspond to the electron energy distributions in Figure \ref{fig-EDacc} are shown in Figure \ref{fig-SEDacc}.  When shock acceleration dominates over both adiabatic expansion and second-order acceleration (red), the spectrum forms narrower emission features.  These correspond to build-up of higher energy particles in the electron distribution. Conversely, as second-order Fermi acceleration becomes more dominant, the spectra broaden, reflecting the smooth broadening of the electron energy distribution.  In both frames of Figure \ref{fig-SEDacc}, the overall spectral flux increases due to the increased energy in the particles, available for radiation, since there was no correction applied to limit the overall energy expenditure in a physical way for this example.

\section{Spectral Analysis of the 3C 279 Flare}

\begin{figure}
\centering
       \includegraphics[width=.9\textwidth]{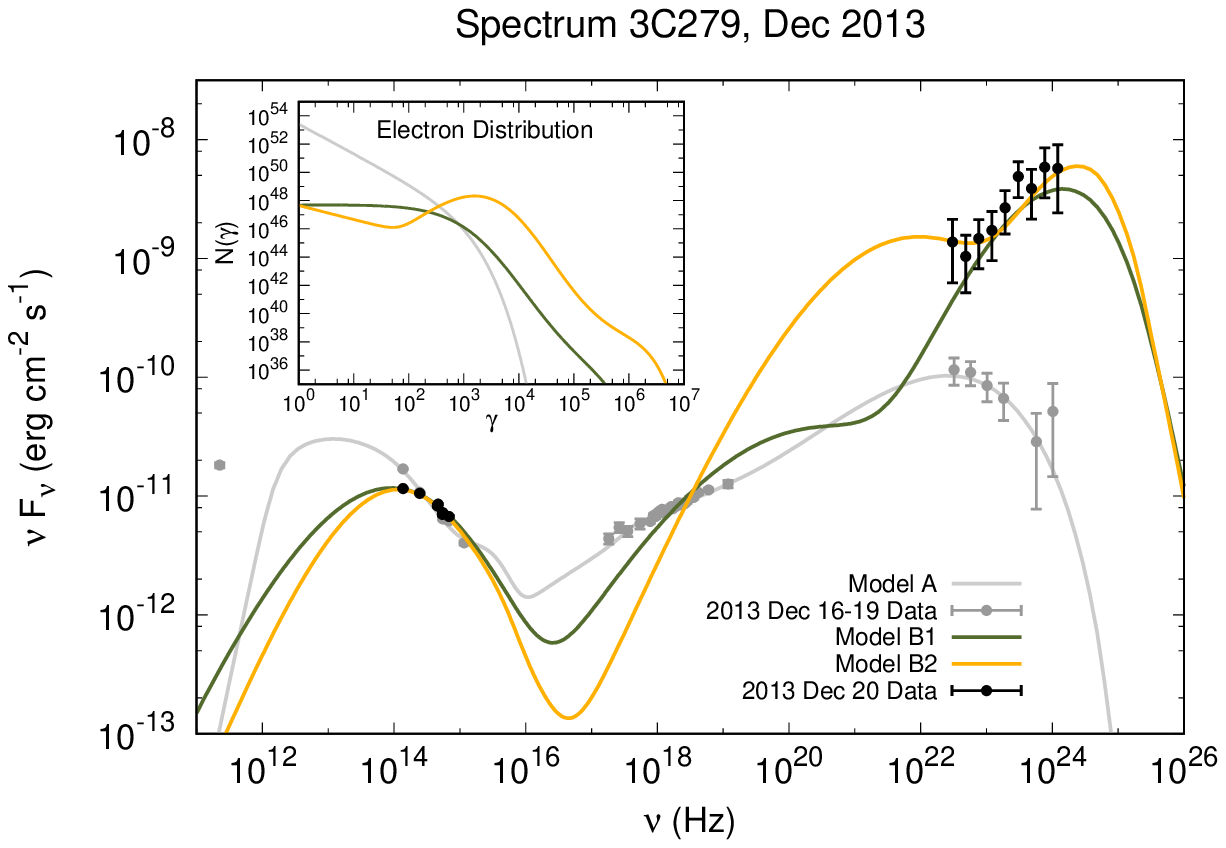}
       \caption{The large panel contains spectral energy distribution simulations with respect to frequency alongside data from 3C 279.  The sub-panel contains the corresponding electron distributions with respect to electron Lorentz factor. The grey data and simulation thereof represent the 3 day quiescent period immediately prior to the flare.  The black data were recorded during the 12-hour $\g$-ray flare period.  The green and yellow curves are were simulated in the numerical model to describe the flare.   All of the components in this figure were published in \cite{lewis19}.  Data from \cite{hayash15}}
       \label{fig-3Cspec}
     \end{figure}

The spectral features from the numerical simulation indicate a relatively strong shock component due to the narrower features (Figure \ref{fig-3Cspec}).  This is echoed in the electron distribution, where the shock build up appears between $\g \approx 10^3$ and $10^4$, especially in yellow (Model B2).  The second order component extends past the shock build up.  The two loss turnovers are due to synchrotron and EC dominating at different energies (see \cite{lewis19} for further details).  The shock build up is less obvious in the green (Model B1) electron distribution because the first-order acceleration is less dominant, and the orange component from Figure \ref{fig-SimpED} is parallel with the x-axis. The second-order acceleration is also not as strong in Model B1. So, Model B2 has more energy overall in both its particles and its emission. 

The following summarizes results from \cite{lewis19}. In other epochs of 3C 279, the electron distribution resembles a power law with an exponential cutoff as expected (see Model A in Figure \ref{fig-3Cspec}). This flare required higher-energy particles, which can be achieved by increasing both first and second-order Fermi acceleration. Additionally, these acceleration mechanisms can act faster than the requisite 2-hour flux doubling timescale for the flare. Thus, magnetic reconnection is unnecessary to explain this flare.  However, reconnection was discussed in the literature of this flare, beginning with the detection paper (\cite{hayash15}), and we therefore discuss it further.  

Those authors estimated the magnetic reconnection of the electron population with observable parameters in an attempt to explain the extreme hardness of the spectrum.  As they explain, magnetic energy can be converted to kinetic energy in the plasma for magnetization $\sigma_B \gtrsim 1$ (e.g. \cite{begelman94,komissarov07}).  In particular, a very hard spectrum can be obtained from simulations wherein the electron magnetization parameter (neglecting protons) $\sigma_e > 100$ (\cite{guo14,sironi14}).  \cite{hayash15} report $\sigma_B \lesssim 10^{-4}$, but argue it is possible (but not that it is observationally verifiable) that the electron magnetization is significantly different from the overall magnetization, implying that reconnection involving electrons could still cause the flare. 

The model described in \cite{lewis19} considers only electrons (and positrons), and does not include a term for magnetic reconnection in the particle transport equation.  However, the model does calculate the maximum Larmor radius for electrons in the emitting region.  In order for the particles to remain inside the jet, their Larmor radius must not exceed the radius of the jet at the emission region location, which is also a result of the model simulations.  The maximum (Larmor limited) electron magnetization parameter $\sigma_{\rm max} \sim$ 2-3.  Since the electron magnetization parameter does not greatly exceed 1, magnetic reconnection from electrons is physically unlikely to make a significant contribution within the framework of this model.  However, the more important result is that first- and second-order Fermi acceleration are sufficient to cause the observed spectrum in the timeframe in which the flare occurred. 

The extreme Compton dominance induced by the isolated $\g$-ray flare, correlates with a sudden decrease in the magnetic field power and thus a departure from equipartition. The lack of simultaneous X-ray observations makes constraining the SSC component during the peak of the flare impossible, and we explored a range of possibilities. However, the energetic requirements of an X-ray flare on par with the $\g$-ray flare are not supported by the available accretion disk power, thus Model B2 is excluded. Energetic particles preferentially occupy larger radii inside the emitting region, and thus may preferentially interact with incident photons from the dust torus and stratified broad line region that impinge on the jet environment.

\bibliographystyle{JHEP}
\bibliography{msNote}

\end{document}